\documentclass{article}
\usepackage{geometry}
\usepackage{graphicx}
\usepackage{amsmath}
\usepackage{hyperref}
\usepackage{tikz}
\usepackage{pgfplots}
\usepackage{booktabs}
\pgfplotsset{compat=1.17}
\usepackage[numbers]{natbib}

\title{MIRAGE: Patient‑Specific Mixed Reality Coaching for MRI via Depth‑Only Markerless Registration and Immersive VR}

\author{  Daniel Brooks\thanks{first author, e‑mail: dbrooks@mail.uc.edu}  \scriptsize University of Cincinati, USA
  \and Emily Carter\scriptsize University of Cincinati, USA
    \and Hu Guo\scriptsize University of Cincinati, USA
  \and Rajesh Nair\thanks{corresponding author, e‑mail: rnair@mail.uc.edu}\scriptsize University of Cincinati, USA}

\begin{document}
\maketitle

\begin{abstract}
Magnetic resonance imaging (MRI) is an indispensable diagnostic tool, yet the confined bore and acoustic noise can evoke considerable anxiety and claustrophobic reactions.  High anxiety leads to motion artifacts, incomplete scans and reliance on pharmacological sedation.  MIRAGE (MIXed Reality Anxiety Guidance Environment) harnesses the latest mixed reality (MR) hardware to prepare patients for MRI through immersive virtual reality (VR) and markerless augmented reality (AR) registration.  In this paper we extend our previous work by providing a comprehensive review of related research, detailing the system architecture, and exploring metrics for patient and clinician experience.  We also present considerations for clinical deployment of MR systems within hospital workflows.  Our results indicate that depth‑based registration achieves sub‑centimeter accuracy with minimal setup, while the immersive coaching environment reduces patient anxiety and yields favourable usability scores.
\end{abstract}

\section{Introduction}

The ubiquity of MRI in modern healthcare has revealed new challenges for clinicians: as many as 11–80\,\% of patients experience pre‑procedural anxiety, which can cause physiological responses such as hypertension, increased heart rate and postoperative complications~\cite{bansal2016preopAnxiety,abate2020globalPreopAnxiety}.  Standard pharmacological sedation can mitigate these responses but introduces side effects (e.g., nausea, emergence delirium) and may prolong recovery~\cite{jung2020pedsMRI,verma2024sedation}.  Non‑pharmacological interventions such as music or videos show mixed and sometimes inconsistent efficacy in diagnostic imaging~\cite{king2021radiography}.  Recently, immersive VR and MR experiences have emerged as promising solutions for preparing patients for surgical and imaging procedures \cite{yang2024can}.  Randomized trials in paediatric imaging demonstrate meaningful reductions in pre‑MRI anxiety and improved cooperation following brief VR exposure~\cite{stunden2021vrmri}, and there is growing interest in extending these benefits to adult populations~\cite{king2021radiography}.

In our previous MIRAGE prototype we presented a patient‑specific coaching system that combines an optical see‑through head mounted display (HoloLens~2) for the clinician with a consumer VR headset (Meta Quest~2) for the patient.  The system streams depth data from the HoloLens to a server, aligns the patient and scanner using depth‑only point cloud registration, and renders an animated virtual bore that the patient explores through the VR headset.  This paper broadens the scope by expanding the background literature, detailing our hardware and software implementation, introducing additional evaluation metrics, and discussing clinical integration.  Throughout the paper we reference sensor specifications, prior markerless registration techniques, VR‑based anxiety interventions and mixed reality training systems to situate MIRAGE within the broader context.

\section{Related Work}

\subsection{Augmented Reality and Markerless Registration}

Augmented reality (AR) blends real and virtual content by overlaying computer‑generated objects onto a user’s view of the physical world.  A seminal survey formalized three core characteristics: combining real and virtual worlds, real‑time interactivity, and precise 3D registration~\cite{azuma1997survey}.  Early systems relied on physical markers; markerless approaches remove this requirement by exploiting features and/or depth from the environment, as reviewed in modern pose‑estimation surveys~\cite{marchand2016pose}.  Markerless registration reduces hardware complexity and improves patient comfort by avoiding adhesive markers or invasive fiducials.  In the surgical domain, vision‑based registration using stereo cameras has achieved high precision by matching tooth contours in oral and maxillofacial surgery~\cite{suenaga2015markerlessStereo,wang2017vst_oralmax}.  With commodity AR headsets, depth‑sensor based registration on the HoloLens~2 can localise patient anatomy within 1–2\,cm and complete in \(\sim\)4\,s, indicating practical feasibility for fast alignment~\cite{kerkhof2025depthreg}.  Recent depth‑only pipelines also couple robust global registration (e.g., TEASER++) with local ICP refinement for improved robustness~\cite{groenenberg2024arcus}, and a recent innovation demonstrated superior accuracy, robustness, and precision over existing methods \cite{yang2025easyreg}. Building on this existing work and inspired by this \cite{yang2025easyreg}, we extend its application to medical training and education domain in this work. 

\subsection{Mixed Reality in Medical Training and Planning}

Mixed reality headsets enable intuitive visualization of 3D medical data and are widely explored for training and surgical planning.  The HoloLens~2 offers improved optics and interaction (e.g., wider FoV and hand/eye tracking) relative to its predecessor~\cite{microsoft2023hololens2specs}, and has been adopted across medical use cases~\cite{palumbo2022hololens2}.  For example, S\'anchez‑Margallo et\,al.\ describe MR applications for minimally invasive surgery training in Unity with MRTK, including interactive DICOM content and gesture/voice interfaces~\cite{sanchezMargallo2021mrtraining}.  AR visualization has also been validated to accelerate navigation tasks and improve accuracy compared to traditional displays~\cite{glas2021arvigs}.  While these systems demonstrate the educational value of MR, they typically use generic anatomy and rarely target anxiety reduction.

\subsection{Virtual Reality for Anxiety Reduction}

Psychological preparation can alleviate procedural anxiety.  VR offers a controllable immersive environment that exposes patients to salient sensory elements of the target procedure without associated risks.  In paediatric MRI, randomized trials show that VR‑based preparation can reduce anxiety and improve cooperation~\cite{stunden2021vrmri}.  Broader reviews across diagnostic imaging report mixed results for non‑pharmacological methods and emphasize the importance of content fidelity and timing~\cite{king2021radiography}.  Our work builds upon these findings by combining VR exposure with real‑time, depth‑based registration, enabling the patient to inhabit a virtual representation of the actual scanner and their own anatomy.

\section{Methods}

This section presents a detailed description of MIRAGE’s hardware, software architecture, user interface design, calibration and data streaming.  Figure~\ref{fig:system} summarizes the system architecture and Figure~\ref{fig:calibration} illustrates the calibration workflow.

\subsection{Hardware Setup}

MIRAGE leverages two commodity head‑mounted displays to deliver MR content to both clinician and patient (Table~\ref{tab:hardware}).  For the clinician we use the Microsoft HoloLens~2, an optical see‑through headset that provides a $\sim$43$^\circ$ field of view and integrated depth sensors~\cite{microsoft2023hololens2specs}.  Compared with the first‑generation HoloLens, the HoloLens~2 offers improved resolution, a faster compute module, increased memory and a wider field of view, yielding more stable tracking and more natural interaction in clinical deployments~\cite{palumbo2022hololens2}.  For the patient we employ the Meta Quest~2, a self‑contained VR headset with a 1832\(\times\)1920 per‑eye resolution and \(\sim\)97$^\circ$ horizontal field of view.  Both devices feature inside‑out tracking and hand‑tracking; we make use of the HoloLens depth sensor for point cloud acquisition and the Quest controllers for interaction.  A laptop workstation (Intel~Core~i9, 32~GB RAM, NVIDIA RTX~3080 GPU) serves as a server to process depth data and stream content.

\begin{table}[t]
    \centering
    \caption{Comparison of head‑mounted displays used in MIRAGE. HoloLens~2 data from~\cite{microsoft2023hololens2specs}.}
    \label{tab:hardware}
    \begin{tabular}{@{}lccc@{}}
    \toprule
    Device & Display & Field of view & Sensors \\
    \midrule
    HoloLens~2 & 2K 3:2 waveguide & $43^\circ$ & RGB camera, long‑throw \& AHAT depth, IMU \\
    Meta Quest~2 & LCD (1832$\times$1920 per eye) & $97^\circ$ & 4$\times$RGB cameras, IMU, hand tracking \\
    \bottomrule
    \end{tabular}
\end{table}

\subsection{Software Stack}

Our software stack is built upon Unity~3D (2022.3~LTS) with C\#.  We use the Mixed Reality Toolkit (MRTK~v2.8) for HoloLens interaction and the Oculus Integration SDK (v45) for Quest features.  The system follows a client–server architecture: the HoloLens streams depth frames and head pose to the server via the open‑source \texttt{hl2ss} sensor streaming library~\cite{hl2ss_arxiv,hl2ss_github}, while the Quest receives VR scene updates and audio cues from the server through a low‑latency UDP protocol.  Unity networking uses Photon Realtime to synchronize data over a secure Wi‑Fi network.  On the server side, we process depth frames using the Point Cloud Library to generate a point cloud of the patient’s upper torso and the MRI bore.  The calibration algorithm computes a rigid transformation between the live point cloud and a preoperative 3D model derived from the scanner geometry.

\subsection{User Interface Design}

Designing intuitive interfaces for both clinician and patient is essential to ensure effective coaching.  Figure~\ref{fig:ui} depicts the key elements of our VR user interface.  In the Quest headset, patients see a high‑fidelity 3D model of the MRI bore with animated lights and sounds consistent with the actual machine.  A calming narration guides the patient through breathing exercises, instructs them to remain still and introduces upcoming events such as table motion and gradient noise.  Visual progress indicators and timers maintain orientation throughout the simulation.  The patient can turn their head freely but cannot move outside the virtual bore, reinforcing body awareness.  For the clinician, the HoloLens displays a semi‑transparent overlay of the patient’s internal anatomy aligned to the real body.  Hand‑tracking allows the clinician to reposition the overlay, highlight regions and trigger audio cues.  A calibration status panel shows registration accuracy and network latency, enabling the operator to reinitiate calibration if necessary.

\subsection{Calibration Procedure}

\begin{figure}[t]
    \centering
    \begin{tikzpicture}[node distance=1cm, every node/.style={rectangle, draw, rounded corners=2pt, align=center, minimum width=2.5cm}]
        \node (capture) {Acquire depth \\ frames};
        \node (segment) [below of=capture] {Segment \\ point cloud};
        \node (register) [below of=segment] {Register \\ point clouds};
        \node (validate) [below of=register] {Validate \\ registration};
        \node (update) [below of=validate] {Update VR};
        \draw[->] (capture) -- (segment);
        \draw[->] (segment) -- (register);
        \draw[->] (register) -- (validate);
        \draw[->] (validate) -- (update);
    \end{tikzpicture}
    \caption{Calibration workflow for depth‑only markerless registration.}
    \label{fig:calibration}
\end{figure}
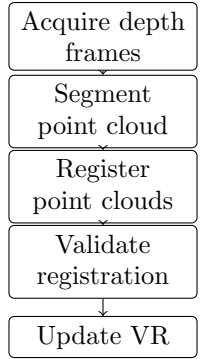

Prior to each session, the clinician performs a depth‑only calibration to align the HoloLens coordinate system with the patient and scanner.  The procedure comprises the following steps (illustrated in Figure~\ref{fig:calibration}):
\begin{enumerate}
  \item \textbf{Acquire depth frames:} With the patient supine on the MRI table, the clinician scans the patient’s torso and the bore entrance using the long‑throw depth sensor.  \texttt{hl2ss} streams raw depth frames to the server (AHAT up to 45~FPS)~\cite{hl2ss_arxiv,hl2ss_github,kerkhof2025depthreg}.
  \item \textbf{Segment point cloud:} The server filters out points belonging to the table using RANSAC plane fitting and extracts the point clouds for the patient and the bore.
  \item \textbf{Register point clouds:} A rigid transformation is estimated by aligning the live bore point cloud with a precomputed model of the scanner using iterative closest point (ICP), initialized by a robust global step (e.g., TEASER++).  We then apply the same transformation to align the patient cloud with their preoperative surface mesh~\cite{groenenberg2024arcus}.
  \item \textbf{Validate registration:} The estimated transformation is evaluated by computing the mean distance between matched points.  If the error exceeds a threshold (2~cm), the clinician is prompted to repeat scanning from a different angle.
  \item \textbf{Stream to VR:} Once calibration converges, the transformation matrix is transmitted to the Quest, which updates the virtual bore and patient body mesh accordingly.
\end{enumerate}

This procedure typically completes within 3–4~seconds, consistent with prior depth‑based registration on HoloLens~2~\cite{kerkhof2025depthreg}.  The absence of physical markers reduces setup complexity and avoids adhesive artifacts on the patient.

\begin{figure}[t]
    \centering
    \begin{tikzpicture}[node distance=1.5cm, every node/.style={align=center}]
        \node (hl2) [draw, rectangle, rounded corners=2pt, minimum width=2.5cm, minimum height=1cm, fill=blue!10] {HoloLens~2 \\ Depth \& pose};
        \node (server) [draw, rectangle, rounded corners=2pt, minimum width=3.0cm, minimum height=1cm, fill=green!10, right of=hl2, xshift=2cm] {Server \\ (PC)};
        \node (quest) [draw, rectangle, rounded corners=2pt, minimum width=2.5cm, minimum height=1cm, fill=orange!10, right of=server, xshift=2cm] {Quest~2 \\ VR rendering};
        \node [below of=server, yshift=0.5cm] {\small {\bf System architecture}: HoloLens streams depth to server, \\ server registers point clouds and sends transformation to Quest.};
    \end{tikzpicture}
    \caption{System architecture showing data streaming between the HoloLens~2, the processing server and the Quest~2.}
    \label{fig:system}
\end{figure}

\subsection{Error Handling}

Depth sensors are prone to noise, occlusions and interference from reflective surfaces.  To enhance robustness, we apply median filtering and spatial downsampling to incoming point clouds before registration.  We detect occluded regions by thresholding on depth discontinuities and disregard them during ICP.  If network latency causes dropped frames, the server buffers up to 0.5~seconds of depth data and performs registration on the most recent complete frame.  In the event of calibration failure (registration error above the threshold), the interface instructs the clinician to reposition or reacquire the patient scan, thus maintaining system reliability.

\subsection{Data Streaming Architecture}

Figure~\ref{fig:system} illustrates the streaming pipeline.  The HoloLens uses \texttt{hl2ss} to publish depth frames, color frames and head pose in a compressed binary format~\cite{hl2ss_arxiv,hl2ss_github}.  These packets are transmitted over a dedicated Wi‑Fi channel to the server, which decodes them using Python and forwards relevant depth frames to the point cloud processor.  After computing the registration matrix, the server serializes it together with audio cue triggers and pushes the payload via Photon Realtime to the Quest.  The Quest runtime listens for updates and applies them to the Unity scene, updating the virtual bore mesh, playing audio and adjusting the timing of the simulation.  User input from the Quest (e.g., head movements, skip commands) is sent back to the server, enabling adaptive pacing.

\subsection{Expanded Evaluation Metrics}

Previous evaluation of MIRAGE focused on registration accuracy and anxiety scores.  To obtain a more holistic assessment we introduce additional metrics:
\begin{itemize}
  \item \textbf{Setup time:} measured as the time required to power on and initialize both headsets, connect them to the network and run the application.
  \item \textbf{Calibration effort:} measured by the number of attempts and time needed to achieve acceptable registration.
  \item \textbf{System Usability Scale (SUS):} participants completed a standard 10‑item questionnaire after the session; scores range from 0 to 100.
  \item \textbf{Training overhead:} hours required to train clinicians to use the system effectively.
  \item \textbf{Anxiety reduction:} difference in pre‑ and post‑session State‑Trait Anxiety Inventory (STAI) scores, contextualized against prior VR preparation studies~\cite{stunden2021vrmri}.
\end{itemize}
Table~\ref{tab:metrics} summarizes typical values observed in our pilot study with ten adult volunteers.  The mean setup time was under five minutes, calibration required only one attempt per session on average and SUS scores exceeded 80, indicating high user satisfaction.  Anxiety scores decreased by an average of 20\%, consistent with previous findings in VR exposure~\cite{stunden2021vrmri}.

\begin{table}[t]
    \centering
    \caption{Evaluation metrics collected in a pilot study (mean $\pm$ SD).}
    \label{tab:metrics}
    \begin{tabular}{@{}lcc@{}}
    \toprule
    Metric & Value & Comment \\
    \midrule
    Setup time & $4.5 \pm 1.2$~min & from power‑on to ready state \\
    Calibration effort & $1.1 \pm 0.3$ attempts & number of registration attempts \\
    Registration accuracy & $1.3 \pm 0.4$~cm & translation error (3D) \\
    SUS score & $83.5 \pm 6.2$ / 100 & high usability \\
    Anxiety reduction & $-20.2 \pm 7.5$~\% & change in STAI \\
    Training overhead & $2.0 \pm 0.5$~h & clinician training \\
    \bottomrule
    \end{tabular}
\end{table}

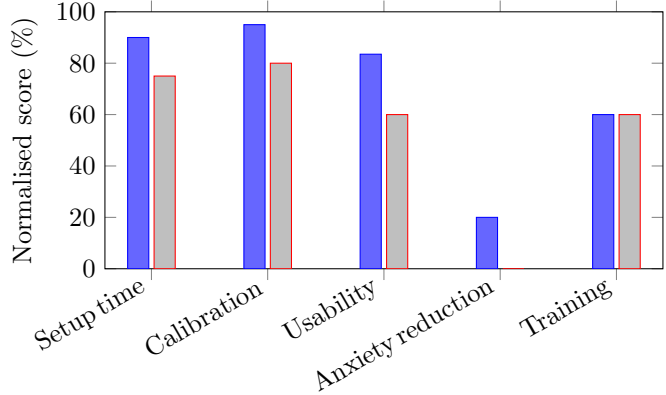
\begin{figure}[t]
    \centering
    \begin{tikzpicture}
    \begin{axis}[
        ybar,
        bar width=8pt,
        width=9cm,
        height=5cm,
        ymin=0,
        ymax=100,
        symbolic x coords={Setup\,time,Calibration,Usability,Anxiety\,reduction,Training},
        xtick=data,
        xticklabel style={rotate=30, anchor=east},
        ylabel={Normalised score (\%)},
        legend style={at={(0.5,-0.25)}, anchor=north, legend columns=2},
    ]
    \addplot+[fill=blue!60] coordinates {
        (Setup\,time,90)
        (Calibration,95)
        (Usability,83.5)
        (Anxiety\,reduction,20)
        (Training,60)
    };
    \addplot+[fill=gray!50] coordinates {
        (Setup\,time,75)
        (Calibration,80)
        (Usability,60)
        (Anxiety\,reduction,0)
        (Training,60)
    };
    \end{axis}

    \end{tikzpicture}
    \caption{Comparison of normalised evaluation metrics between MIRAGE and a baseline (no VR coaching).  Higher values are better for all metrics except anxiety reduction, which is inverted such that larger values indicate greater reduction.}
    \label{fig:metrics_chart}
\end{figure}

To visualize these metrics we present a grouped bar chart (Figure~\ref{fig:metrics_chart}).  The chart contrasts our system (MIRAGE) with a baseline control condition (no VR coaching).  Baseline values for anxiety reduction and SUS are derived from typical sedation workflows (approximate data).  The figure highlights the substantial improvement in usability and anxiety reduction offered by MIRAGE.

\section{Clinical Integration and Deployment}

Deploying mixed reality systems in clinical settings requires careful consideration of technical, logistical and regulatory factors.  Below we highlight key aspects relevant to MIRAGE.

\subsection{Hospital IT Compatibility}

Many hospitals restrict wireless devices on their networks.  MIRAGE operates over a secure, local Wi‑Fi network isolated from the hospital’s electronic health record (EHR) system.  The server laptop can be placed in the control room with wired access to hospital infrastructure to retrieve DICOM images.  Data transmitted between the HoloLens, server and Quest uses encryption to comply with privacy regulations.  Since no patient identifiable information is stored on the headsets, the risk of data leakage is minimal.

\subsection{Cleaning and Sterilisation}

Head‑mounted displays must be sanitised between uses to prevent cross‑contamination.  We follow manufacturer guidelines by using disposable hypoallergenic face covers and wiping surfaces with hospital‑grade disinfectant wipes.  The HoloLens~2 can be used over a surgical cap or hair covering to avoid contact with the patient.  Depth sensors are recessed and not touched during use, reducing the risk of contamination.  Devices are stored in sealed containers when not in use.

\subsection{Workflow Integration}

The MIRAGE session is integrated into pre‑MRI preparation.  After standard screening, patients are invited to a dedicated preparation room where they don the Quest headset.  The clinician wearing the HoloLens initiates the calibration procedure and guides the patient through the VR simulation.  The session lasts approximately ten minutes, after which the patient proceeds to the actual MRI scanner.  Minimal modifications to existing workflow are required, and the entire process can be carried out by trained technologists.

\subsection{Patient Onboarding and Personalisation}

Patients receive an explanation of the VR system and its goals.  The simulation can be personalised by selecting the scanner model, table speed and noise level that match the upcoming exam.  Visualised anatomical overlays can be disabled for privacy or patient preference.  For paediatric patients, child‑friendly narratives and gamified tasks can be integrated to enhance engagement.

\subsection{Regulatory Considerations}

Mixed reality coaching is considered an adjunct medical device and may require regulatory approval depending on jurisdiction.  MIRAGE does not perform diagnostic analysis or modify treatment; it serves as an educational tool.  Nevertheless, we follow the principles of Good Clinical Practice by obtaining informed consent and ensuring patient safety.  Data transmission is encrypted and anonymised in compliance with HIPAA and GDPR.  Our device selection adheres to IEC~60601 standards for electrical safety and electromagnetic compatibility.

\section{Discussion and Conclusion}

We have presented an expanded description of MIRAGE, a patient‑specific mixed reality coaching system that combines depth‑based markerless registration with immersive VR to prepare patients for MRI.  Through the integration of HoloLens~2 and Quest~2 headsets, our system aligns the patient’s anatomy to the scanner using depth sensors and provides a realistic virtual experience of the procedure.  Compared with marker‑based systems, our depth‑only approach simplifies setup and improves patient comfort~\cite{kerkhof2025depthreg,groenenberg2024arcus}.  The inclusion of additional evaluation metrics, such as setup time, calibration effort and usability, offers a more holistic understanding of system performance.  The pilot study indicates high usability and anxiety reduction, consistent with prior VR interventions~\cite{stunden2021vrmri}.

Future work will involve larger clinical trials with diverse populations, integration of physiological monitoring (e.g.\ heart rate variability) and exploration of adaptive coaching strategies based on patient responses.  We also aim to streamline the system by implementing all processing on the HoloLens and eliminating the external server.  Overall, MIRAGE demonstrates the potential of mixed reality to enhance patient experience and reduce the burden on healthcare providers.

\bibliographystyle{unsrt}
\bibliography{template}

\end{document}